\documentclass[aps,pra,showpacs,twocolumn]{revtex4-1}
\usepackage{amsmath}
\usepackage{amssymb}
\usepackage{graphicx}
\usepackage{dcolumn}
\usepackage{bm}

\begin{document}

\title{Bell states and entanglement of two-dimensional polar molecules in
electric fields}
\author{Ying-Yen Liao}
\email{yyliao@nuk.edu.tw}
\affiliation{Department of Applied Physics, National University of Kaohsiung, Kaohsiung
811, Taiwan}
\date{\today }

\begin{abstract}
Entanglement generated from polar molecules of two-dimensional rotation is
investigated in a static electric field. The electric field modulates the
rotational properties of molecules, leading to distinctive entanglement. The
concurrence is used to estimate the degree of entanglement. When the
electric field is applied parallel or perpendicular to the intermolecular
direction, the concurrences reveal two overlapping features. Such a
pronounced signature corresponds to the coexistence of all Bell-like states.
The characteristics of Bell-like states and overlapping concurrences are
kept independent of the modulation of dipole-field and dipole-dipole
interactions. On the contrary, the Bell-like states fail to coexist in other
field directions, reflecting nonoverlapping concurrences. Furthermore, the
thermal effect on the entanglement is analyzed for the Bell-like states.
Dissimilar suppressed concurrences occur due to different energy structures
for the two specific field directions.
\end{abstract}

\maketitle

\section{Introduction}

Exploring entanglement has attracted considerable interest in various
physical systems, ranging from atoms and photons to solid-state materials
\cite{Wilk07,Weber09,Neumann08,Shulman12}. Entanglement describes a
phenomenon in which two or more subsystems are linked through direct or
indirect interactions. This a\ quantum phenomenon has become an essential
ingredient in quantum information processing \cite{Amico08,Horodecki09}.
Among various architectures, a bipartite system is a fundamental unit to
exploit the nature of entanglement. Fascinating properties have been
extracted from a variety of bipartite systems. The Bell state is a
representative example, which plays an important role in dense coding
protocol \cite{Bennett92}, quantum teleportation \cite{Bennett93}, and
entanglement swapping \cite{Zukowski93}. As a result, generating
entanglement becomes an indispensable step for follow-up operations and
further applications.

Two polar molecules can form a platform for generating entanglement directly
through dipole-dipole interaction. This molecular platform provides an
advantage of controllable rotational properties via the coupling of external
fields with the dipole moment \cite{DeMille02,Karra16,Han16}. Field sources
include static electric fields \cite{Wei11,Wei112}, laser pulses \cite%
{Charron07}, and optimized complex pulses \cite{Mishima09}. The rotational
properties, such as energy and wave function, are modified to affect the
entanglement as well as the dipole-dipole interaction. In most studies,
polar molecules\ are considered to rotate in three dimensions. However, only
minor efforts have been made that focus on the entanglement generated by
two-dimensional rotation. The energies, level degeneracy, and wave functions
of\ a two-dimensional rotor are different from those of a three-dimensional
rotor \cite{McIntyre12}. Two-dimensional rotational manner has been observed
in the experiments of hydrogen on a surface \cite%
{Svensson99,Bengtsson00,Billy02}. The planar rotation in diverse
configurations further leads to interesting consequences for energy spectrum
\cite{Landman84,Shih03}, molecular orientation \cite{Shima04}, and specific
heat \cite{Shima05,Iwata10}. In this work we study the entanglement in a
system of two polar molecules confined to a plane. Both molecules rotate in
two dimensions, which are modified by applying an in-plane electric field.
We use concurrence to qualify the degree of entanglement by orienting the
electric field. For specific field directions, all Bell-like states are
jointly displayed, corresponding to a situation of overlapping concurrences.
Specifically, Bell-like states are robustly maintained when the dipole-field
and dipole-dipole interactions are tuned. On the contrary, these pronounced
properties are destroyed in an electric field of another direction. The
obtained results are different from those in the three-dimensional cases. We
further analyze the thermal effect on the entanglement for the Bell-like
states.
\begin{figure}[tbh]
\includegraphics[width=8.5cm]{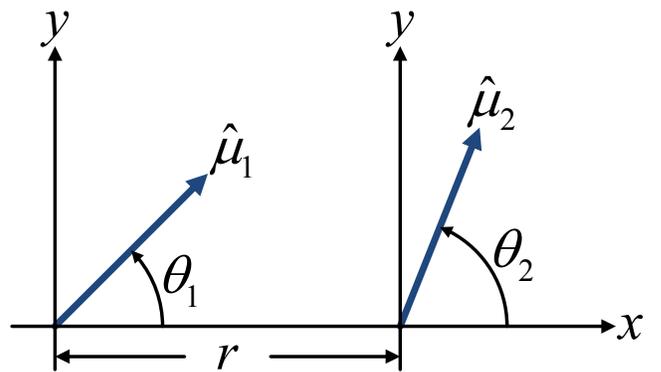}
\caption{Schematic diagram of two polar molecules separated by a distance $r$%
. The\ left and right molecules are confined to two-dimensional motion,
corresponding to the unit vectors of the dipole moments $\hat{\protect\mu}%
_{1}=(\cos \protect\theta _{1},\sin \protect\theta _{1})$ and $\hat{\protect%
\mu}_{2}=(\cos \protect\theta _{2},\sin \protect\theta _{2})$, respectively.
}
\label{fig1}
\end{figure}

\section{Model and theory}

We consider two polar diatomic molecules of rotational constant $B$ and
dipole moment $\mu $ confined to a plane as depicted in Fig. \ref{fig1}.
Both molecules rotate in two dimensions, which are spatially separated by an
interval $r$. A static electric field of strength $\varepsilon $ and tilt
angle $\theta _{t}$ is applied to couple with the dipole moments of the
molecules. The modified two-dimensional rotation leads to specific
dipole-dipole interaction. The Hamiltonian describing the system is
\begin{equation}
H=\sum_{i=1}^{2}H_{0}^{i}+V_{d},  \label{hamiltonian}
\end{equation}%
with%
\begin{equation}
H_{0}^{i}=-B\frac{d^{2}}{d\theta _{i}^{2}}-\omega (\cos \theta _{t}\cos
\theta _{i}+\sin \theta _{t}\sin \theta _{i}),  \label{H0}
\end{equation}%
\begin{equation}
V_{d}=\Omega \left[ \widehat{\mu }_{1}\cdot \widehat{\mu }_{2}-3\left(
\widehat{\mu }_{1}\cdot \widehat{u}_{x}\right) \left( \widehat{\mu }%
_{2}\cdot \widehat{u}_{x}\right) \right] ,  \label{ddinteraction}
\end{equation}%
where $\omega =\mu \varepsilon $ is the strength of dipole-field
interaction, $\Omega =\mu ^{2}/4\pi \epsilon _{0}r^{3}$ is the strength of
dipole-dipole interaction, $\widehat{\mu }_{i}=(\cos \theta _{i},\sin \theta
_{i})$ is the unit vector of the dipole moment, $\widehat{u}_{x}=(1,0)$ is
the unit vector, $\epsilon _{0}$ is the permittivity of free space, and $%
\theta _{i}$ is the polar angle between the molecular axis and the $x$ axis
for the left $(i=1)$ and right $(i=2)$ molecules. From the coordinates of
the molecules, the dipole-dipole interaction $V_{d}$ is given by

\begin{equation}
V_{d}=\Omega \left( \sin \theta _{1}\sin \theta _{2}-2\cos \theta _{1}\cos
\theta _{2}\right) .  \label{dipole}
\end{equation}%
The magnitudes of $\omega $ and $\Omega $ in Eqs. (\ref{H0}) and (\ref%
{dipole}) are dependent on the dipole moment, field strength, and
intermolecular distance.

For a single molecule in an electric field, the field-dependent energy $%
\varepsilon _{l}$ and its corresponding wave function $\psi _{l}$ are
obtained from the equation
\begin{equation}
H_{0}\psi _{l}=\varepsilon _{l}\psi _{l},  \label{sho}
\end{equation}%
where the index $i$ is omitted and the positive integer $l$ is defined from $%
0$. The wave function $\psi _{l}$ is described by
\begin{equation}
\psi _{l}(\theta )=\sum_{m}c_{m}^{l}\varphi _{m}(\theta ),  \label{who}
\end{equation}%
where $c_{m}^{l}$ is the expanded coefficient, $\varphi _{m}(\theta )=\exp
(im\theta )/\sqrt{2\pi }$ is the field-free eigenfunction, corresponding to
the energy $m^{2}B$ for $m=0,\pm 1,\pm 2,\cdots $ \cite{McIntyre12}.

The effect of the dipole-dipole interaction on the system is captured by
employing the lowest two states of the molecules. The chosen states are the
ground state $\left\vert 0_{i}\right\rangle =\left\vert \psi _{0}^{i}(\theta
_{i})\right\rangle $ and the first excited state $\left\vert
1_{i}\right\rangle =\left\vert \psi _{1}^{i}(\theta _{i})\right\rangle $,
corresponding to the energies $\varepsilon _{0}$ and $\varepsilon _{1}$,
respectively. Here the index $i$ is restored. To solve the Hamiltonian in
Eq. (\ref{hamiltonian}), a composite state is taken by the form\
\begin{equation}
\left\vert \Psi _{n}\right\rangle =d_{1}^{n}\left\vert 00\right\rangle
+d_{2}^{n}\left\vert 01\right\rangle +d_{3}^{n}\left\vert 10\right\rangle
+d_{4}^{n}\left\vert 11\right\rangle ,  \label{wavefunction}
\end{equation}%
where $d_{j}^{n}$ is the coefficient, and $\left\vert 00\right\rangle $ $%
=\left\vert 0_{1}\right\rangle \otimes \left\vert 0_{2}\right\rangle $, $%
\left\vert 01\right\rangle $ $=\left\vert 0_{1}\right\rangle \otimes
\left\vert 1_{2}\right\rangle $, $\left\vert 10\right\rangle $ $=\left\vert
1_{1}\right\rangle \otimes \left\vert 0_{2}\right\rangle $, and $\left\vert
11\right\rangle $ $=\left\vert 1_{1}\right\rangle \otimes \left\vert
1_{2}\right\rangle $ are the basis states. By using $\left\vert \Psi
_{n}\right\rangle $, the energy $E_{n}$ and the coefficient $d_{j}^{n}$ can
be derived from the matrix
\begin{equation}
\widetilde{H}=\left(
\begin{array}{cccc}
\delta _{0,0}+\Gamma _{0,0} & \Gamma _{0,X} & \Gamma _{0,X} & \Gamma _{X,X}
\\
\Gamma _{0,X}^{\ast } & \delta _{0,1}+\Gamma _{0,1} & \Gamma _{X,XC} &
\Gamma _{1,X} \\
\Gamma _{0,X}^{\ast } & \Gamma _{X,XC} & \delta _{0,1}+\Gamma _{0,1} &
\Gamma _{1,X} \\
\Gamma _{X,X}^{\ast } & \Gamma _{1,X}^{\ast } & \Gamma _{1,X}^{\ast } &
\delta _{1,1}+\Gamma _{1,1}%
\end{array}%
\right) ,  \label{matrix}
\end{equation}%
with $\delta _{\alpha ,\beta }=\varepsilon _{\alpha }+\varepsilon _{\beta }$
and $\Gamma _{\alpha ,\beta }=\Omega (S_{\alpha }S_{\beta }-2C_{\alpha
}C_{\beta })$ for $\alpha $ and $\beta $ being $0$, $1$, $X$, and $XC$. The
factors are expressed as $C_{0(1)}=\left\langle \psi _{0(1)}^{i}\left\vert
\cos \theta _{i}\right\vert \psi _{0(1)}^{i}\right\rangle $, $%
S_{0(1)}=\left\langle \psi _{0(1)}^{i}\left\vert \sin \theta _{i}\right\vert
\psi _{0(1)}^{i}\right\rangle $, $C_{X(XC)}=\left\langle \psi
_{0(1)}^{i}\left\vert \cos \theta _{i}\right\vert \psi
_{1(0)}^{i}\right\rangle $, and $S_{X(XC)}=\left\langle \psi
_{0(1)}^{i}\left\vert \sin \theta _{i}\right\vert \psi
_{1(0)}^{i}\right\rangle $.

The density matrix method is used to examine the entanglement properties in
the system \cite{Wootters98,Arnesen01,Wang01}. Based on ${{E_{n}}}$ and $%
\left\vert \Psi _{n}\right\rangle $, the density matrix can be described as
\begin{equation}
\rho =\frac{{\sum\limits_{n=1}^{4}{\exp \left( {-{E_{n}}/{k_{B}}T}\right)
\left\vert \Psi _{n}\right\rangle \langle \Psi _{n}|}}}{{\sum%
\limits_{n=1}^{4}{\exp \left( {-{E_{n}}/{k_{B}}T}\right) }}},
\label{density}
\end{equation}%
with the Boltzmann constant $k_{B}$\ and temperature $T$. The degree of
entanglement can be measured by the concurrence
\begin{equation}
C=\max \left\{ 0,\sqrt{\lambda _{1}}-\sqrt{\lambda _{2}}-\sqrt{\lambda _{3}}-%
\sqrt{\lambda _{4}}\right\} ,  \label{concurrence}
\end{equation}%
where the quantities $\lambda _{k}$ are the eigenvalues of the matrix
\begin{equation}
\widetilde{\rho }=\rho \left( \sigma _{y}\otimes \sigma _{y}\right) \rho
^{\ast }\left( \sigma _{y}\otimes \sigma _{y}\right) ,  \label{r_matrix}
\end{equation}%
in decreasing order. The operator $\sigma _{y}$ is the Pauli matrix and $%
\rho ^{\ast }$ is the complex conjugate of $\rho $. The value of the
concurrence ranges from zero to one, corresponding to an unentangled state
and a maximally entangled state, respectively. $\ $
\begin{figure}[th]
\includegraphics[width=8.5cm]{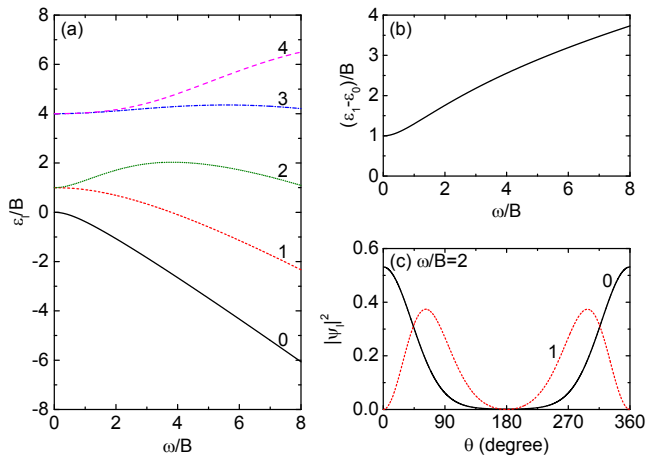}
\caption{(a) Energies $\protect\varepsilon _{l}/B$ as a function of the
parameter $\protect\omega /B$. The energy difference between the ground and
first excited states is shown in (b). (c) Probability densities $\left\vert
\protect\psi _{l}\right\vert ^{2}$ of the ground and first excited states as
a function of the angle $\protect\theta $ for $\protect\omega /B=2$. The
tilt angle is $\protect\theta _{t}=0^{\text{o}}$.}
\label{fig2}
\end{figure}

\section{Numerical results and discussion}

In this work the systemic properties are explored by controlling the
strength of the dipole-field interaction $\omega $ and the dipole-dipole
interaction $\Omega $. The energy-related quantities are expressed in units
of $B$ throughout the calculations. Figure \ref{fig2}(a) shows the energy
spectra of a single molecule in an electric field of $\theta _{t}=0^{\text{o}%
}$, i.e., the field direction lies along the $x$ axis. The energy properties
are influenced by varying the parameter $\omega /B$. When the field is
absent, all excited states are two-fold degenerate except for the ground
state at $\omega /B=0$. Due to the presence of dipole-field interaction, the
degenerate levels are split off at $\omega /B\neq 0$. The energy structure
differs from that of a three-dimensional rotor \cite{Meyenn70,Rost92}. For
the lowest two states, the energies $\varepsilon _{0}$ and $\varepsilon _{1}$
display decreasing behavior with increasing the parameter $\omega /B$. The
energy difference between the two levels is gradually increased from $B$
[see Fig. \ref{fig2}(b)]. For instance, a difference of $1.75B$ is
determined at $\omega /B=2$. Correspondingly, in Fig. \ref{fig2}(c) the
probability densities of the two states are modulated to spread along the $x$
axis. As a whole, the energies are the same at arbitrary $\theta _{t}$,
whereas the spatial distributions of the wave functions depend on the field
direction. It is worth noting that, as defined in Eq. (\ref{wavefunction}),
a two-level subsystem is built from the two individual states $\left\vert
\psi _{0}\right\rangle $ and $\left\vert \psi _{1}\right\rangle $. However,
the two states $\left\vert \psi _{1}\right\rangle $ and $\left\vert \psi
_{2}\right\rangle $ are degenerate at $\omega /B=0$, and thus have the same
energies $\varepsilon _{1}=\varepsilon _{2}=B$ \cite{McIntyre12}. The
corresponding wave function is either $\exp (i\theta )/\sqrt{2\pi }$ or $%
\exp (-i\theta )/\sqrt{2\pi }$. At $\omega /B=0$, the transition between the
states $\left\vert \psi _{0}\right\rangle $ and $\left\vert \psi
_{2}\right\rangle $ can occur as well as the transition between the states $%
\left\vert \psi _{0}\right\rangle $ and $\left\vert \psi _{1}\right\rangle $%
. The two-level subsystem fails in this situation. As a result, to generate
entanglement based on coupled two-level subsystems, the tuning of $\omega
/B=0$ is excluded throughout the calculation.
\begin{figure}[th]
\includegraphics[width=8.5cm]{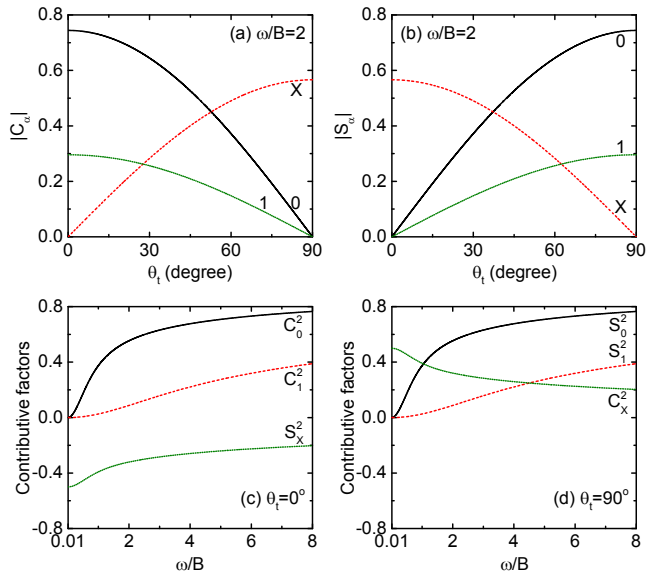}
\caption{Absolute values $\left\vert C_{\protect\alpha }\right\vert $ (a)
and $\left\vert S_{\protect\alpha }\right\vert $ (b) of the factors as a
function of the tilt angle $\protect\theta _{t}$ for $\protect\omega /B=2$.
Square values $C_{\protect\alpha }^{2}$ (c) and $S_{\protect\alpha }^{2}$
(d) of the contributive factors as a function of the parameter $\protect%
\omega /B$ for $\protect\theta _{t}=0^{\text{o}}$ and $90^{\text{o}}$. }
\label{fig3}
\end{figure}

In Eq. (\ref{matrix}) the energy ${{E_{n}}}$ and the state ${{\left\vert
\Psi _{n}\right\rangle }}$ depend on these factors. The properties of the
wave function plays a determinant role in the contribution of the factors.
In Figs. \ref{fig3}(a) and (b) an electric field of fixed strength is
oriented to modulate the spatial distribution of the wave function so that
the absolute values of the factors become angle-dependent. By increasing $%
\theta _{t}$, the\ value $\left\vert C_{X}\right\vert $ inversely varies
against $\left\vert C_{0}\right\vert $ and $\left\vert C_{1}\right\vert $.
The features of $\left\vert S_{\alpha }\right\vert $ are similar to those of
$\left\vert C_{\alpha }\right\vert $. If the tilt angle is $\theta _{t}=0^{%
\text{o}}$ or $90^{\text{o}}$, the values of the factors will be either real
or purely imaginary. For $\theta _{t}=0^{\text{o}}$, only $\left\vert
C_{0}\right\vert $, $\left\vert C_{1}\right\vert $, and $\left\vert
S_{X}\right\vert $ are contributive and maximal, whereas $\left\vert
S_{0}\right\vert $, $\left\vert S_{1}\right\vert $, and $\left\vert
C_{X}\right\vert $ are zero. A reverse situation occurs for $\theta _{t}=90^{%
\text{o}}$ where the contributive roles are interchanged to be $\left\vert
S_{0}\right\vert $, $\left\vert S_{1}\right\vert $, and $\left\vert
C_{X}\right\vert $. To probe the two cases, the square values of the
contributive factors are shown in Figs. \ref{fig3}(c) and (d). We simply set
the initial value of $\omega /B$ as $0.01$ to avoid the three-level
situation mentioned above. The relations $C_{0}^{2}=S_{0}^{2}$, $%
C_{1}^{2}=S_{1}^{2}$, and $C_{X}^{2}=-S_{X}^{2}$ hold at arbitrary $\omega
/B $. The magnitudes of $C_{0}^{2}$ and $C_{1}^{2}$, as well as those of\ $%
S_{0}^{2}$ and $S_{1}^{2}$, are raised by increasing the parameter $\omega
/B $. On the contrary, $C_{X}^{2}$ and $S_{X}^{2}$ are close to zero at
large $\omega /B$. \
\begin{figure}[th]
\includegraphics[width=8.5cm]{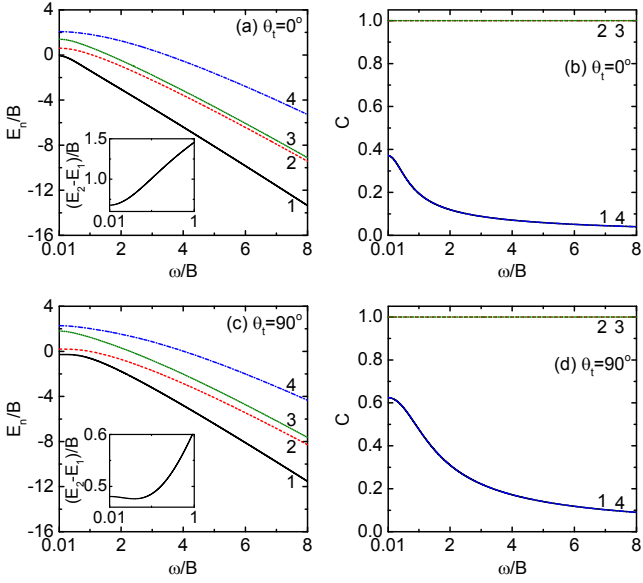}
\caption{Energies $E_{n}/B$ (a) and concurrences $C$ (b) as a function of
the parameter $\protect\omega /B$ for $\protect\theta _{t}=0^{\text{o}}$.
(c) and (d) correspond to $\protect\theta _{t}=90^{\text{o}}$. The insets
show the difference between the energies $E_{1}$ and $E_{2}$. The fixed
parameter is $\Omega /B=0.8$.}
\label{fig4}
\end{figure}

The entanglement property is analyzed in an electric field of a specific
tilt angle. Figure \ref{fig4}(a) shows the four energy levels for $\theta
_{t}=0^{\text{o}}$ and $\Omega /B=0.8$. In this situation the field
direction is parallel to the intermolecular direction. The energies are
gradually diminished by increasing $\omega /B$. The corresponding
concurrences reveal pronounced features as illustrated in Fig. \ref{fig4}%
(b). The concurrence for $\left\vert \Psi _{1}\right\rangle $ is equal to
that for $\left\vert \Psi _{4}\right\rangle $. Both values are decreased by
increasing $\omega /B$. By contrast, the states $\left\vert \Psi
_{2}\right\rangle $ and $\left\vert \Psi _{3}\right\rangle $ have the
maximum concurrences $C=1$ all the time. When the electric field is applied
perpendicular to the intermolecular direction, the decreasing energies in
Fig. \ref{fig4}(c) are similar to those for $\theta _{t}=0^{\text{o}}$.
However, the energy structures are different in the two cases. For example,
the difference between $E_{1}$ and $E_{2}$ for $\theta _{t}=0^{\text{o}}$ is
greater than\ that for $\theta _{t}=90^{\text{o}}$. Different from the case
of $\theta _{t}=0^{\text{o}}$, a local minimum for $\theta _{t}=90^{\text{o}%
} $ is observed at $\omega /B\simeq 0.299$, i.e., an anticrossing occurs
[see insets of Fig. \ref{fig4}]. In Fig. \ref{fig4}(d) the overlapping
features of concurrences also occur in the case of $\theta _{t}=90^{\text{o}%
} $. The concurrences for $\left\vert \Psi _{2}\right\rangle $ and $%
\left\vert \Psi _{3}\right\rangle $ are still $C=1$, while the same
concurrences for $\left\vert \Psi _{1}\right\rangle $ and $\left\vert \Psi
_{4}\right\rangle $ are higher than those in Fig. \ref{fig4}(b).
\begin{figure}[th]
\includegraphics[width=8.5cm]{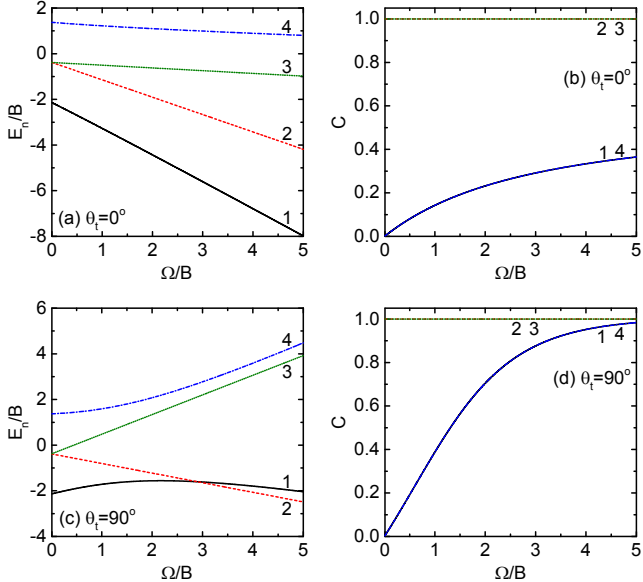}
\caption{Energies $E_{n}/B$ (a) and concurrences $C$ (b) as a function of
the parameter $\Omega /B$ for $\protect\theta _{t}=0^{\text{o}}$. (c) and
(d) correspond to $\protect\theta _{t}=90^{\text{o}}$. The fixed parameter
is $\protect\omega /B=2$.}
\label{fig5}
\end{figure}

Next, we turn to examining the effect of dipole-dipole interaction on the
energies and concurrences by\ fixing the strength of the dipole-field
interaction. The fixed parameter is set to be $\omega /B=2$, corresponding
to $\varepsilon _{0}=-1.07B$ and $\varepsilon _{1}=0.68B$ in Fig. \ref{fig2}%
. By increasing the parameter $\Omega /B$, the four levels for $\theta
_{t}=0^{\text{o}}$ reveal decreasing features [see Fig. \ref{fig5}(a)]. The
energy differences between the adjacent levels increase at large $\Omega /B$%
. However, in Fig. \ref{fig5}(c) a diverse energy structure is obtained for $%
\theta _{t}=90^{\text{o}}$. The levels for $E_{1}$ and $E_{2}$ have a
crossing at $\Omega /B\simeq 2.9$. Increasing behavior is observed for $%
E_{3} $ and $E_{4}$. As shown in Figs. \ref{fig5}(b) and (d), these
concurrences can be classified into two types, which are similar to those in
Fig. \ref{fig4}. Such a result indicates that the overlapping property is
robust against the modulation of dipole-dipole interaction. For a system of
three-dimensional rotors, similar concurrences can be observed only when the
parameter is $\omega /B=0$ \cite{Wei11}. Furthermore, the stronger
dipole-dipole interaction obviously enhances the concurrences for $%
\left\vert \Psi _{1}\right\rangle $ and $\left\vert \Psi _{4}\right\rangle $%
. The degree of enhancement on the concurrence is higher in the case of $%
\theta _{t}=90^{\text{o}}$.

To gain insight into the distinctive entanglement properties, the matrix in
Eq. (\ref{matrix}) can be simplified according to the results in Fig. \ref%
{fig3}:
\begin{equation}
\widetilde{H}=\left(
\begin{array}{cccc}
2\varepsilon _{0}+a & 0 & 0 & d \\
0 & \varepsilon _{0}+\varepsilon _{1}+b & f & 0 \\
0 & f & \varepsilon _{0}+\varepsilon _{1}+b & 0 \\
d & 0 & 0 & 2\varepsilon _{1}+c%
\end{array}%
\right) ,  \label{reducedmatrix}
\end{equation}%
where the parameters $a=-2\Omega C_{0}^{2}$, $b=-2\Omega C_{0}C_{1}$, $%
c=-2\Omega C_{1}^{2}$, $d=\Omega S_{X}^{2}$, and $f=\Omega \left\vert
S_{X}\right\vert ^{2}$ are for $\theta _{t}=0^{\text{o}}$ while $a=\Omega
S_{0}^{2}$, $b=\Omega S_{0}S_{1}$, $c=\Omega S_{1}^{2}$, $d=-2\Omega
C_{X}^{2}$, and $f=-2\Omega \left\vert C_{X}\right\vert ^{2}$ are for $%
\theta _{t}=90^{\text{o}}$. After calculating the matrix, one obtains the
Bell-like states
\begin{equation}
\left\vert \Psi _{1}\right\rangle =\frac{\Delta +\sqrt{d^{2}+\Delta ^{2}}}{%
\sqrt{2N_{+}}}\left\vert 00\right\rangle -\frac{d}{\sqrt{2N_{+}}}\left\vert
11\right\rangle ,  \label{Bell1s}
\end{equation}%
\begin{equation}
\left\vert \Psi _{2(3)}\right\rangle =\frac{1}{\sqrt{2}}\left\vert
01\right\rangle -\frac{1}{\sqrt{2}}\left\vert 10\right\rangle ,
\label{Bell2s}
\end{equation}%
\begin{equation}
\left\vert \Psi _{3(2)}\right\rangle =\frac{1}{\sqrt{2}}\left\vert
01\right\rangle +\frac{1}{\sqrt{2}}\left\vert 10\right\rangle ,
\label{Bell3s}
\end{equation}%
\begin{equation}
\left\vert \Psi _{4}\right\rangle =\frac{\Delta -\sqrt{d^{2}+\Delta ^{2}}}{%
\sqrt{2N_{-}}}\left\vert 00\right\rangle -\frac{d}{\sqrt{2N_{-}}}\left\vert
11\right\rangle ,  \label{Bell4s}
\end{equation}%
and their energies%
\begin{equation}
E_{1}=\varepsilon _{0}+\varepsilon _{1}+(a+c)/2-\sqrt{d^{2}+\Delta ^{2}},
\label{Bell1e}
\end{equation}%
\begin{equation}
E_{2(3)}=\varepsilon _{0}+\varepsilon _{1}+b-f,  \label{Bell2e}
\end{equation}%
\begin{equation}
E_{3(2)}=\varepsilon _{0}+\varepsilon _{1}+b+f,  \label{Bell3e}
\end{equation}%
\begin{equation}
E_{4}=\varepsilon _{0}+\varepsilon _{1}+(a+c)/2+\sqrt{d^{2}+\Delta ^{2}},
\label{Bell4e}
\end{equation}%
with $\Delta =\varepsilon _{1}-\varepsilon _{0}+(c-a)/2$, $%
N_{+}=d^{2}+\Delta ^{2}+\Delta \sqrt{d^{2}+\Delta ^{2}}$, and $%
N_{-}=d^{2}+\Delta ^{2}-\Delta \sqrt{d^{2}+\Delta ^{2}}$. For $\theta
_{t}=0^{\text{o}}$, the subscripts $2$ are used in Eqs. (\ref{Bell2s}) and (%
\ref{Bell2e}), corresponding to $3$ used in Eqs. (\ref{Bell3s}) and (\ref%
{Bell3e}). The subscripts interchange for $\theta _{t}=90^{\text{o}}$. The
concurrence is also derived from the definition
\begin{equation}
C=2\left\vert d_{2}^{n}d_{3}^{n}-d_{1}^{n}d_{4}^{n}\right\vert ,  \label{yuc}
\end{equation}%
for the state $\left\vert \Psi _{n}\right\rangle $ \cite{Yu2002}. The
concurrences for $\left\vert \Psi _{2}\right\rangle $ and $\left\vert \Psi
_{3}\right\rangle $ are straightforwardly $C=1$. The states $\left\vert \Psi
_{1}\right\rangle $ and $\left\vert \Psi _{4}\right\rangle $ have the same
concurrence
\begin{equation}
C=\frac{1}{\sqrt{1+\Delta ^{2}/d^{2}}},  \label{BellC}
\end{equation}%
with
\begin{equation}
\frac{\Delta ^{2}}{d^{2}}=\left[ \frac{\varepsilon _{1}-\varepsilon
_{0}-\Omega (C_{1}^{2}-C_{0}^{2})}{\Omega S_{X}^{2}}\right] ^{2},  \label{r0}
\end{equation}%
for $\theta _{t}=0^{\text{o}}$ and
\begin{equation}
\frac{\Delta ^{2}}{d^{2}}=\left[ \frac{\varepsilon _{1}-\varepsilon
_{0}+\Omega (S_{1}^{2}-S_{0}^{2})/2}{2\Omega C_{X}^{2}}\right] ^{2},
\label{r90}
\end{equation}%
for $\theta _{t}=90^{\text{o}}$. The analytic results obtained agree exactly
with the energies and concurrences in Figs. \ref{fig4} and \ref{fig5}. Once
the particular field directions are selected, the four Bell-like states can
jointly occur as shown in Eqs. (\ref{Bell1s}-\ref{Bell4s}). This situation
corresponds to two pairs of identical concurrences. More importantly, the
modulation of the dipole-field and dipole-dipole interactions does not
destroy the coexistent characteristic of Bell-like states.
\begin{figure}[th]
\includegraphics[width=8.5cm]{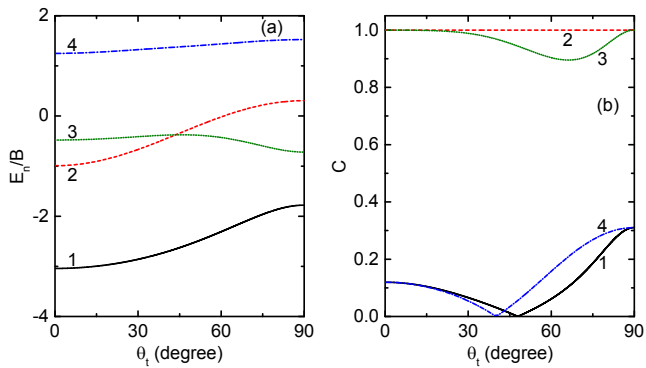}
\caption{Energies $E_{n}/B$ (a) and concurrences $C$ (b) as a function of
the tilt angle $\protect\theta _{t}$ for $\protect\omega /B=2$ and $\Omega
/B=0.8$.}
\label{fig6}
\end{figure}

When the tilt angle is not limited to $\theta _{t}=0^{\text{o}}$ and $90^{%
\text{o}}$, all factors become contributive to energy and concurrence.
Figure \ref{fig6}(a) shows the angle-dependent energies for $\omega /B=2$
and $\Omega /B=0.8$. Here the levels are labeled based on the case of $%
\theta _{t}=0^{\text{o}}$; this label is consistently retained throughout
the tuning of $\theta _{t}$. The energies $E_{1}$ and $E_{4}$ vary
monotonically with increasing $\theta _{t}$. By contrast, the energies $%
E_{2} $ and $E_{3}$ are modulated to intersect at $\theta _{t}\simeq 43.5^{%
\text{o}}$. For $\theta _{t}\neq 0^{\text{o}}$ and $90^{\text{o}}$, an
explicit signature of entanglement is that the concurrences no longer
coincide together [see Fig. \ref{fig6}(b)]. Only the state $\left\vert \Psi
_{2}\right\rangle $ retains the maximum concurrence $C=1$, corresponding to $%
(\left\vert 01\right\rangle -\left\vert 10\right\rangle )/\sqrt{2}$
throughout the modulation of $\theta _{t}$. The other states are composed of
the four basis states in Eq. (\ref{wavefunction}), whose concurrences are
smaller than $1$. The result implies that the four Bell-like states do not
jointly exist at $\theta _{t}\neq 0^{\text{o}}$ and $90^{\text{o}}$.
Specifically, due to the angle-dependency of the factors, the concurrences
for $\left\vert \Psi _{1}\right\rangle $ and $\left\vert \Psi
_{4}\right\rangle $ are modulated to be greatly reduced at $\theta
_{t}\simeq 47.9$ and $40$, respectively.
\begin{figure}[th]
\includegraphics[width=8.5cm]{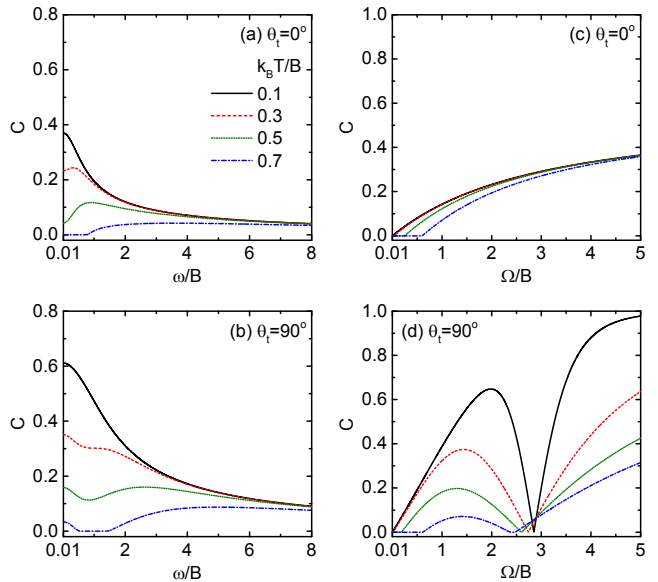}
\caption{Concurrences $C$ for $\protect\theta _{t}=0^{\text{o}}$ and $90^{%
\text{o}}$ as a function of the parameters $\protect\omega /B$ and $\Omega
/B $ at different temperatures. (a) and (b) correspond to $\Omega /B=0.8$
while $\protect\omega /B=2$ is used in (c) and (d).}
\label{fig7}
\end{figure}

The thermal influence on the entanglement is analyzed, specifically with
respect to the Bell-like states. Figures \ref{fig7}(a) and (b) show the
temperature-related concurrences by varying the parameter $\omega /B$. Since
the temperature induces the contribution of the four Bell-like states, the
mixing effect leads to a suppressed concurrence [see Eq. (\ref{density})].
When the temperature is raised, the degree of reduction of concurrence is
enlarged. Although the concurrences for $\theta _{t}=0^{\text{o}}$ and $90^{%
\text{o}}$ are similar, two dissimilar suppressed features occur, especially
in the region of small $\omega /B$. This result originates from the two
different energy structures, as illustrated in Figs. \ref{fig4}(a) and (c).
Furthermore, the properties of the energy structure strongly affect the
concurrences that are dependent on the parameter $\Omega /B$ [see Figs. \ref%
{fig7}(c) and (d)]. Different from smooth features for $\theta _{t}=0^{\text{%
o}}$, greatly suppressed concurrences can be obtained around the point where
an energy crossing between $E_{1}$ and $E_{2}$ occurs for $\theta _{t}=90^{%
\text{o}}$, as depicted in Figs. \ref{fig5}(a) and (c).

\section{Conclusions}

We have investigated the entanglement properties in a system of
two-dimensional polar molecules by applying an electric field. The electric
field parameters modulate the two-dimensional rotation, leading to
distinctive energy spectra and concurrence. When the field direction is
perpendicular or parallel to the intermolecular direction, the concurrences
become overlapping. The four Bell-like states are collectively generated in
the two situations. The coexistence of Bell-like states steadily holds
against the modulation of the dipole-field and dipole-dipole interactions.
Once other field directions are selected, these significant properties will
be destroyed. Instead, the concurrences can be greatly reduced in certain
field directions. The concurrences for the Bell-like states are further
analyzed at finite temperatures. The different energy structures cause
diverse suppressed features in the two situations.

\begin{acknowledgments}
This work is supported by the Ministry of Science and Technology of Taiwan
under Grant No. 103-2112-M-390-003-MY3.
\end{acknowledgments}

\end{document}